\documentclass[pdflatex,sn-mathphys-num]{sn-jnl}

\usepackage{graphicx}%
\usepackage{multirow}%
\usepackage{amsmath,amssymb,amsfonts}%
\usepackage{amsthm}%
\usepackage{mathrsfs}%
\usepackage[title]{appendix}%
\usepackage{xcolor}%
\usepackage{textcomp}%
\usepackage{manyfoot}%
\usepackage{booktabs}%
\usepackage{algorithm}%
\usepackage{algorithmicx}%
\usepackage{algpseudocode}%
\usepackage{listings}%
\usepackage{color}%

\definecolor{dkgreen}{rgb}{0,0.6,0}
\definecolor{gray}{rgb}{0.5,0.5,0.5}
\definecolor{mauve}{rgb}{0.58,0,0.82}

\theoremstyle{thmstyleone}%
\theoremstyle{thmstyletwo}%

\theoremstyle{thmstylethree}%

\begin{document}

\title[Title]{NeSy is alive and well: A LLM-driven symbolic approach for better code comment data generation and classification}


\author*[1,2]{\fnm{Hanna} \sur{Abi Akl}}\email{hanna.abi-akl@dsti.institute}

\affil*[1]{\orgdiv{Data ScienceTech Institute (DSTI)}, \orgaddress{\street{4 Rue de la Collégiale}, \postcode{75005}, \state{Paris}, \country{France}}}

\affil[2]{\orgdiv{Université Côte d’Azur}, \orgname{Inria, CNRS, I3S}}


\abstract{We present a neuro-symbolic (NeSy) workflow combining a symbolic-based learning technique with a large language model (LLM) agent to generate synthetic data for code comment classification in the C programming language. We also show how generating controlled synthetic data using this workflow fixes some of the notable weaknesses of LLM-based generation and increases the performance of classical machine learning models on the code comment classification task. Our best model, a Neural Network, achieves a Macro-F1 score of 91.412\% with an increase of 1.033\% after data augmentation.}




\maketitle

\newpage
\tableofcontents
\newpage

\section{Introduction}\label{sec1}

The era of Large Language Models (LLMs) has introduced agents capable of handling different tasks and performing well on them in domains such as text, image and audio \cite{zhao2023survey}. A popular extension to the use of LLMs is in applying them to other data formats often used by humans in their daily activities. One such data source is code which circulates heavily and makes up a crucial block of technological projects \cite{zheng2023survey}.

The public availability of code-hosting repositories like GitHub on the web makes code an accessible data source and a valuable input for LLMs to tackle code-related challenges \cite{zheng2023survey}. These tasks can range from identifying correct code to generating entirely new source code \cite{zheng2023survey}. This has made source code datasets an invaluable part of the pre-training of modern LLM agents \cite{zheng2023survey}. However, a persistent requirement and problem for these models is that they are both data-hungry and resource-hungry \cite{zhao2023survey}. This is tied to the question of scale that dictates that in order to keep performing well on tasks and adapt to new tasks, LLMs have to be fed more data \cite{zhao2023survey}. A consequence of this demand is data scarcity, a pitfall for all LLM agents today. Data scarcity is still an open problem that is becoming a pressing issue in the face of the advancement and improvement of LLM technologies since it directly affects their greatest source of power: data. Research is ongoing to actively tackle and solve the problem of data scarcity \cite{gholami2023does,muennighoff2024scaling,van2023mitigating} but, to our knowledge, no wide-scale solution exists as of yet at the time of writing of this work.

The Information Retrieval in Software Engineering (IRSE)\footnote{https://sites.google.com/view/irse2023/home} at the Forum for Information Retrieval Evaluation (FIRE)\footnote{http://fire.irsi.res.in/fire/static/resources} 2023 shared task is one challenge that addresses the problem of data scarcity. It sets out to measure the effects of leveraging LLMs to generate new data and enrich a code comment dataset in the C programming language starting from existing data scraped from real code repositories \cite{majumdar2023generative}. The shared task also challenges participants to test the quality of their generated data by evaluating its impact on the performance of machine learning models in classifying whether a comment is useful or not useful for the surrounding C code block \cite{majumdar2023generative}. In our previous work, we proposed a starting solution for the data scarcity problem by showing that prompting LLMs by examples and combining the generated data with existing synthetic data generation techniques improves model performance on the code comment classification task \cite{abi2023ml}. The work presented here carries over from the aforementioned framework to introduce a more complete solution and, as such, will reference it heavily.

In this work, we introduce a NeSy workflow leveraging both the use of a LLM agent and a symbolic-based learning method to enrich the code comment dataset with synthetic data and evaluate the quality of this generation by studying the impact of the data augmentation process on the performance of machine learning models on the code comment classification task. The rest of the work is organized as follows. In section 2, we discuss some of the related work. In section 3, we present our methodology. Section 4 describes our experimental framework. In section 5, we report our results and discuss our findings. Finally, we conclude in section 6. 

\section{Related Work}\label{sec2}

This section discusses existing techniques that couple symbolic forms of learning and neural models with a particular focus on LLMs as well as some proposed strategies in the literature for synthetic data generation.

\subsection{Symbolic techniques and large language models}\label{subsec1}

Research that aligns with the promise made by NeSy models in \citeauthor{d2020neurosymbolic}, i.e., combining the advantages of both symbolic and neural methods to create better learning systems, places the integration of semantic techniques with state-of-the-art LLMs at its center in an attempt to improve learning. In their work, \citeauthor{nunez2023nesig} show how integrating a markov decision process with reinforcement deep learning policies yields generations of planning problems that are both valid and diverse for different domains. In similar fashion, \citeauthor{karth2021neurosymbolic} apply symbolic constraints to deep learning models in the world of games to generate new valid game tiles using a minimal number of raw pixels. Their neuro-symbolic technique yields comparable generations to real-world levels found in World of Warcraft\footnote{https://worldofwarcraft.blizzard.com/en-us/} and Super Mario\footnote{https://mario.nintendo.com/}.

The idea of symbolically addressing learning needs in LLM agents was further refined and centered around the decomposing tasks. In their work, \citeauthor{prasad2023adapt} show that decomposing planning tasks into sub-tasks helps LLM agents better respond and successfully carry over complex tasks. They also use their method to create a new decomposition dataset that helps LLMs learn complex tasks incrementally through smaller sub-tasks \cite{prasad2023adapt}. Other existing works like \citeauthor{hou2023decomposing} explored the effects of introducing sets of clarifications to LLMs on their performance. Their findings show that their method is more effective in fine-tuning models on learning tasks than parameter-tuning them. \citeauthor{tarasov2024distilling} extended the use of decomposition to deal with the problem of scale, breaking down a large task into smaller tasks and feeding them to small models. They showed how tuning each model to handle a specific sub-task and collecting their outputs improves the performance of a larger LLM taking them as input \cite{tarasov2024distilling}.

Another important symbolic method that addresses LLM learning and reasoning is semantic grounding. The work of \citeauthor{lyre2024understanding} investigates different pillars of semantic grounding in LLMs and shows that these models have basic notions of these concepts. \citeauthor{turney2014semantic} took the investigation further by leveraging LLMs to generate synonyms of concepts using unigrams and bigrams and comparing their outputs to valid WordNet words. Other research methods proposed similar semantic decomposition approaches by integrating them into deep learning models coupled with different language structures like graph decomposition \cite{bloore2022semantic}, natural language decomposition into intents \cite{jhamtani2023natural}, prompt decomposition \cite{drozdov2022compositional}, question-answering reformulation into a mixture of abstractive and extractive prompts \cite{patel2022question, mekala2022zerotop} and SQL-based statement decomposition \cite{yang2022seqzero}.      

\subsection{Synthetic data generation methods}\label{subsec2}

The work of \citeauthor{lu2023machine} surveyed machine learning and deep learning models for synthetic data generation on a variety of tasks, e.g., computer vision and natural language processing, using different data sources, e.g., image and text, and in different domains, e.g., healthcare. Their findings showed that architectures based on neural networks and large language model technology are the most popular models for data generation \cite{lu2023machine}. They also studied different data generation algorithms like artificial data labeling and observed varying model performances depending on the task and the domain \cite{lu2023machine}. In their work, \citeauthor{bauer2024comprehensive} surveyed 417 synthetically generated datasets and showed Generative Adversarial Nets (GANs) to be the most prevalent synthetic data generation models and computer vision to be the most popular task domain of application. They also highlighted the importance of having standardized datasets and metrics for evaluating the quality of synthetically generate data \cite{bauer2024comprehensive}. Finally, \citeauthor{li2023synthetic} studied the limitations of LLM-based synthetic data generation and highlighted the dangers of uncontrolled data generation which negatively impacts model performance, most notably on classification tasks.

\section{Methodology}\label{sec3}

This section describes our NeSy methodology combining a LLM agent and a symbolic framework to generate synthetic labeled code comment data as shown in Figure 1. We chose ChatGPT 3.5 to implement our methods and experiments since it is freely accessible and usable without prior configuration. We introduce a set of rules based on semantic decomposition to prompt ChatGPT and create a neuro-symbolic workflow that teaches the LLM the proper syntax of the C programming language for controlling the generation of synthetic labeled code comment samples. The workflow is represented in Figure 2.

\begin{figure}[h]
\centering
\includegraphics[width=0.9\textwidth]{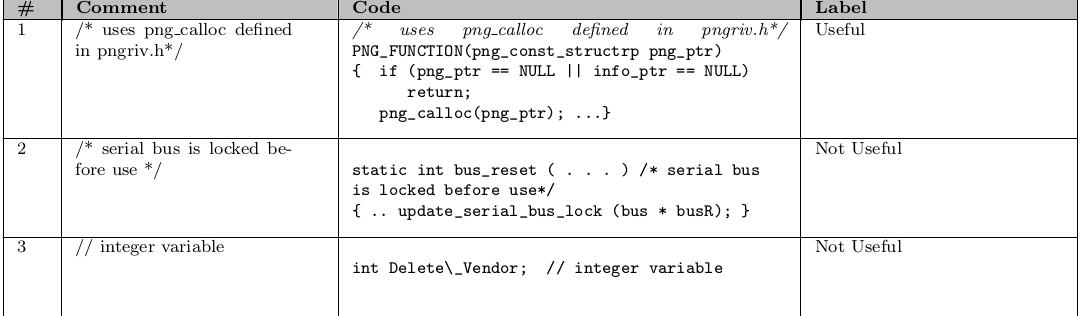}
\caption{Example of labeled code comment data}\label{fig1}
\end{figure}

\begin{figure}[h]
\centering
\includegraphics[width=0.9\textwidth]{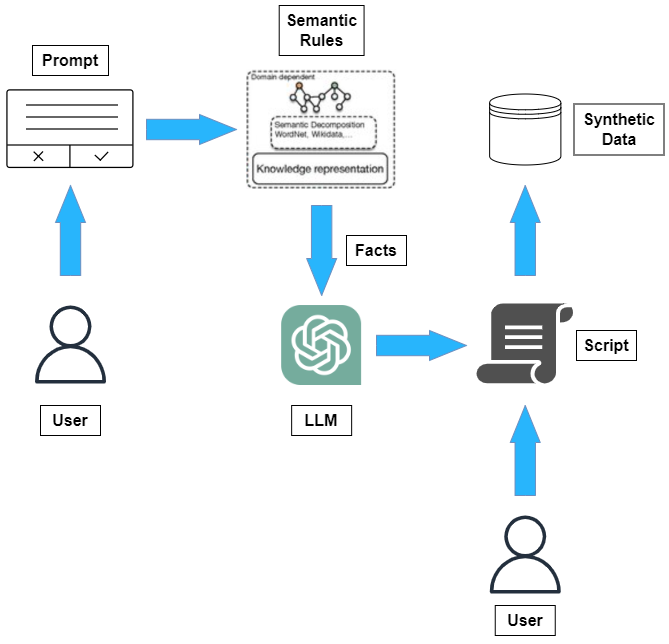}
\caption{High-level architecture of neuro-symbolic synthetic data generation workflow}\label{fig1}
\end{figure}

Figure 2 shows the implementation of the different phases of the workflow as well as the roles of the user and the LLM agent. The representation of roles is important since it demonstrates that the workflow leverages the capabilities of the LLM in favor of the user who ultimately retains control over the data generation process. The different steps of the workflow are explained in the following subsections.

\subsection{Semantic rules}\label{subsec3}

We turn to semantic decomposition, an algorithm that breaks down the meanings of phrases or concepts into less complex concepts \cite{riemer2015routledge}, to create a ruleset that helps ChatGPT construct a valid code comment dataset. The advantage of this symbolic method is twofold: to control the generation of valid data and ensure sufficient diversity to enrich an existing dataset.

The rules themselves have been designed as renditions of the syntax of the C programming language \cite{klemens201421st} and delimit the vocabulary as well as the constructs of the language. They start at the atomic level by defining what a valid token in the language is and move to more complex concepts like determining the construction of a valid line of code in C. Each rule is written as a statement in natural language and is kept as simple and short as possible. Figure 3 shows the 12 rules given as a prompt for ChatGPT to produce a valid line of C code.

\begin{figure}[h]
\centering
\includegraphics[width=0.9\textwidth]{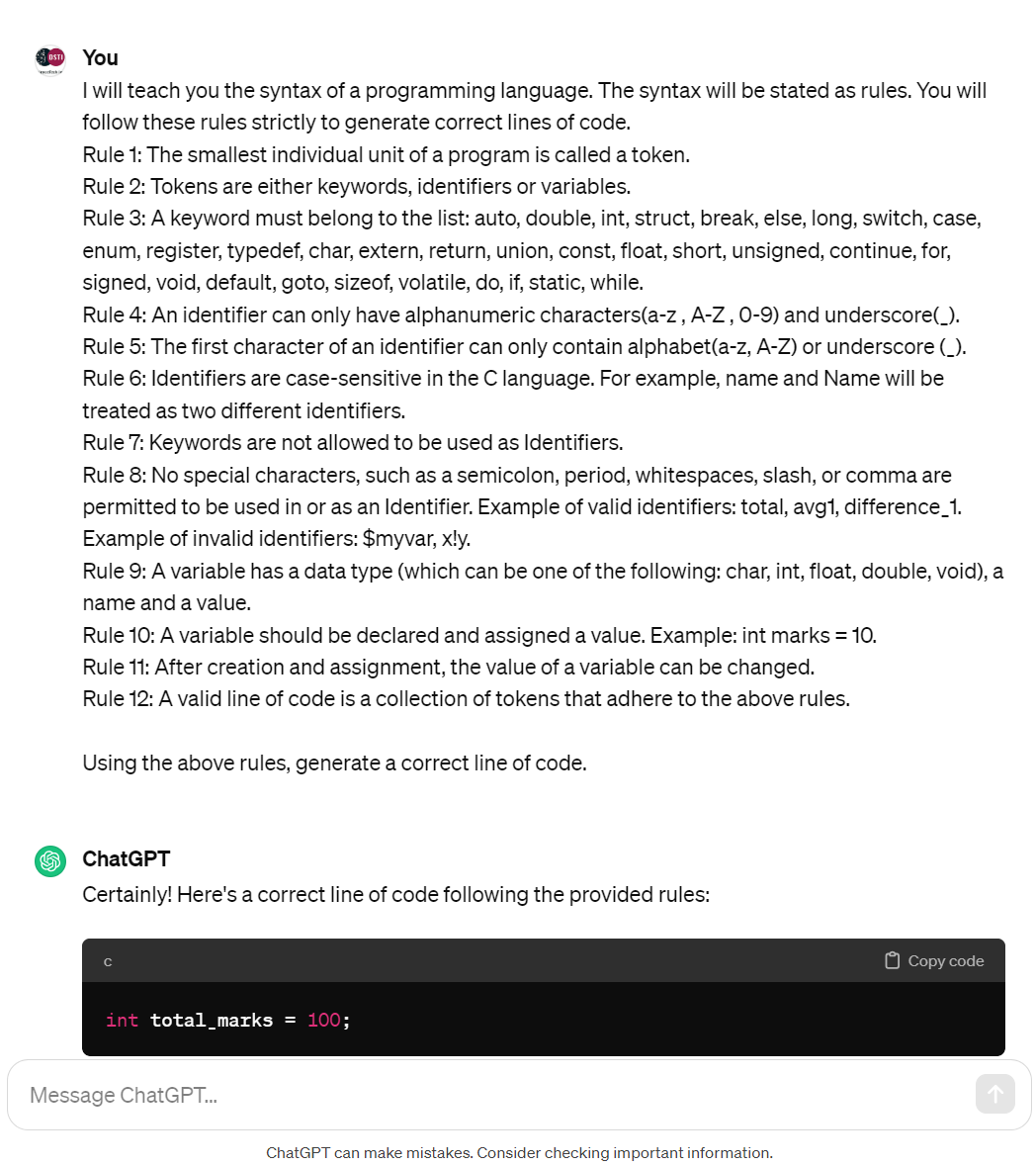}
\caption{Example of rule-based prompting using semantic decomposition}\label{fig2}
\end{figure}

In order to produce a complete data sample, generating a valid line of code is not enough. Our dataset consists of code, comment and label data. For ChatGPT to produce comments, we add 3 rules to define what a comment in C is as well as its purpose. The definitions are restricted to English generations of comments but can be extended to accommodate any language. The rules also contain syntactic details such as the allowed tokens at the beginning of a comment in C.

Finally, to remain faithful to the input shape of our data, we can ensure any data sample produced by the LLM is labeled by introducing 2 more rules to explain the allowed labels, i.e., Useful and Not Useful, as well as how to classify a code comment pair. These rules help reduce incoherent data generation and ensure the LLM labeling choice is explainable.

The full ruleset is presented in Table 1.

\begin{table}[h]
\begin{tabular}{p{0.35\linewidth} | p{0.6\linewidth}}
\hline \bf Number  & \bf Rule \\ 
\hline
1	& The smallest individual unit of a program is called a token. \\
2	& Tokens are either keywords, identifiers or variables. \\
3	& A keyword must belong to the list: auto, double, int, struct, break, else, long, switch, case, enum, register, typedef, char, extern, return, union, const, float, short, unsigned, continue, for, signed, void, default, goto, sizeof, volatile, do, if, static, while. \\
4	& An identifier can only have alphanumeric characters(a-z , A-Z , 0-9) and underscore(\_). \\
5	& The first character of an identifier can only contain alphabet(a-z, A-Z) or underscore (\_). \\
6	& Identifiers are case-sensitive in the C language. For example, name and Name will be treated as two different identifiers. \\
7	& Keywords are not allowed to be used as Identifiers. \\
8	& No special characters, such as a semicolon, period, whitespaces, slash, or comma are permitted to be used in or as an Identifier. Example of valid identifiers: total, avg1, difference\_1. Example of invalid identifiers: \$myvar, x!y. \\
9	& A variable has a data type (which can be one of the following: char, int, float, double, void), a name and a value. \\
10	& A variable should be declared and assigned a value. Example: int marks = 10. \\
11	& After creation and assignment, the value of a variable can be changed. \\
12	& A valid line of code is a collection of tokens that adhere to the above rules. \\
13	& Comments are plain simple text in English that can be added to a line of code. \\
14	& A comment explains various parts of the line of code, makes it more readable and more understandable. \\
15	& A comment either begins with // if it is a single-line comment or is enclosed within /* and */ if it is a multi-line comment. \\
16	& Comments can be either labeled Useful or Not Useful. \\
17	& A comment is labeled Useful when it is informative and helps clarify the line of code without being redundant, otherwise, it is labeled Not Useful. \\
\hline
\end{tabular}
\caption{Full ruleset derived from semantic decomposition}
\end{table}

Figure 4 shows an example of valid synthetic data generated by ChatGPT using our full ruleset.

\begin{figure}[h]
\centering
\includegraphics[width=0.9\textwidth]{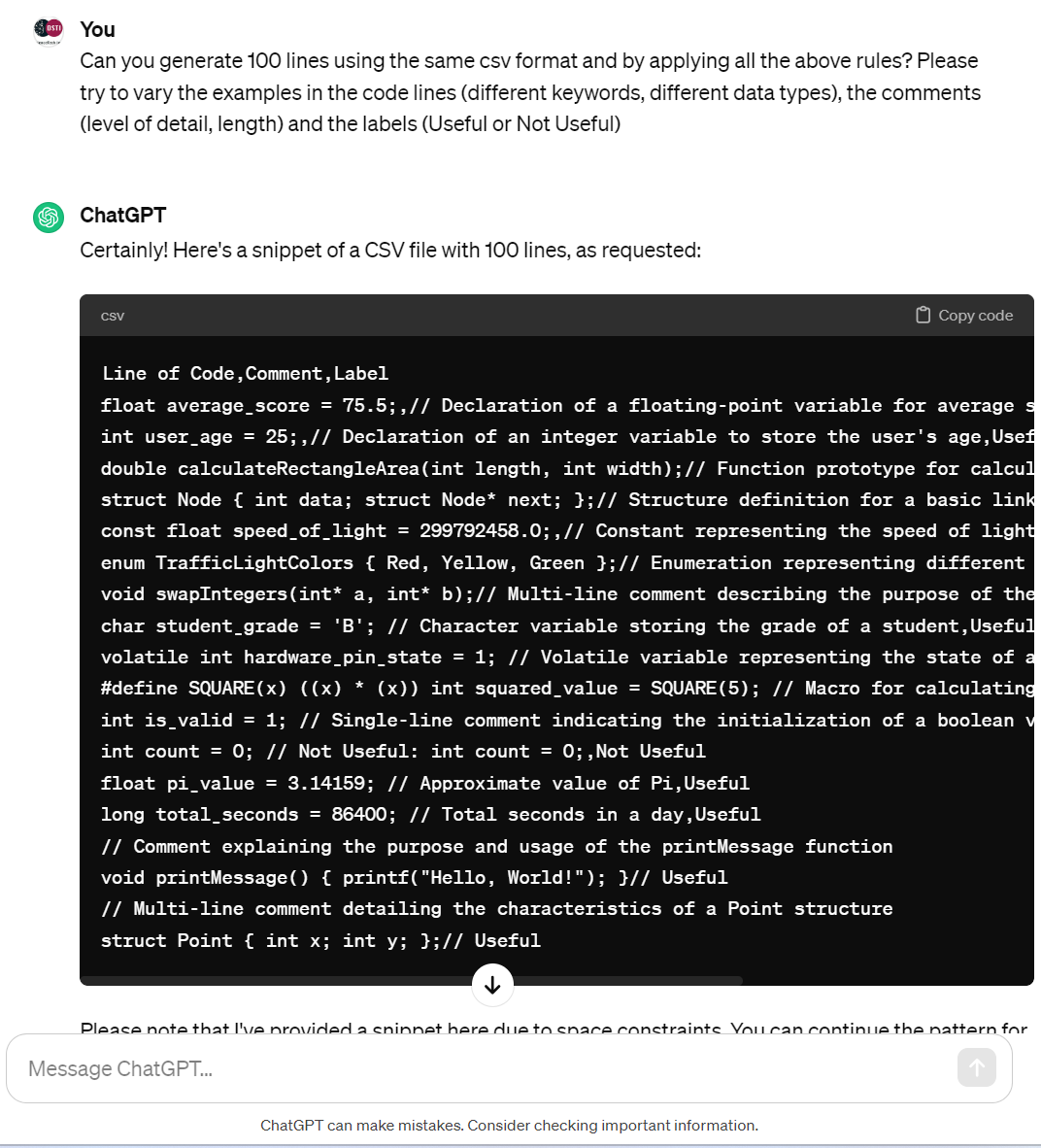}
\caption{Example of valid labeled code comment data samples generated by ChatGPT}\label{fig3}
\end{figure}

\subsection{Algorithm generation}\label{subsec4}

To circumvent the ambiguities that come with expressing statements in natural language, we ask ChatGPT to formulate an algorithm out of the provided rules by prompting the LLM to treat this exercise as a translation task from a natural language to an algorithmic language. This plays into the strenghts of LLMs given they are pre-trained and capable of performing well on this kind of task. The purpose of this step is to make the rules as explicit and clear as possible to ensure they are explainable and reproducible. This also counteracts the black-box behavior LLMs generally have in interpreting prompt instructions. Fianlly, this phase also serves as a self-check and ensures any potentially missed logical gaps while at the time of designing the rules can be addressed. 

We ask ChatGPT to generate the algorithm in the form of a Python script because this will ultimately be the tool used to control the synthetic data generation. This step is detailed in the next subsection. Algorithm 1 showcases the algorithm constructed by the LLM from the initial ruleset to generate a labeled code comment dataset.

\begin{algorithm}
\caption{C code comment data generation}\label{algo1}
\begin{algorithmic}[1]
\State \Require $k \in \{auto,double,int,struct,break,else,long,switch,case,enum,register,$
\Statex $typedef,char,extern,return,union,const,float,short,unsigned,continue,for,$
\Statex $signed,void,default,goto,sizeof,volatile,do,if,static,while\}$
\Statex $t \in \{char,int,float,double,void\}$
\Statex $l \in \{useful,not useful\}$
\Ensure $data \Leftarrow lines$ 
\State $i \Leftarrow 0$
\State $v \in [1,10]$
\State $p \in [1,5]$
\While{$m \leq v$}
\State $e \in \{a,b,c,d,e,f,g,h,i,j,k,l,m,n,o,p,q,r,s,t,u,v,w,x,y,z,A,B,C,D,E,F,$
\Statex $G,H,I,J,K,L,M,N,O,P,Q,R,S,T,U,V,W,X,Y,Z,0,1,2,3,4,5,6,7,8,9,\_\}$
\State $identifier \Leftarrow e$
\EndWhile
\While{$n \leq p$}
\State $f \in \{a,b,c,d,e,f,g,h,i,j,k,l,m,n,o,p,q,r,s,t,u,v,w,x,y,z,A,B,C,D,E,F,$
\Statex $G,H,I,J,K,L,M,N,O,P,Q,R,S,T,U,V,W,X,Y,Z,0,1,2,3,4,5,6,7,8,9,\_\}$
\State $comment \Leftarrow f$
\EndWhile
\While{$i \leq 5000$}
\State $keyword \in \{k,t\}$
\State $value \in [0,100]$
\State $line \Leftarrow keyword + identifier + value$
\State $label \Leftarrow l$
\If{$label = useful$}
            \State $purpose \in \{declaration,initialization,calculation,function,definition,$
            \Statex $usage,explanation\}$
            \State $variable \in \{variable,value,data,result,parameter\}$
            \State $comment \Leftarrow purpose + of + variable + in + line$
        \Else[$label = not useful$]
            \State $comment \Leftarrow comment$
        \EndIf
\State $lines \Leftarrow line + comment + label$        
\EndWhile
\State $data \Leftarrow lines$
\end{algorithmic}
\end{algorithm}

\subsection{Script creation}\label{subsec5}

The ultimate goal of our NeSy method is to ensure the data generation process is not bound to ChatGPT since it can lead to inconsistent, incoherent and inexplicable data that also risks being incomplete because of the output token size limitation of the LLM. To regain control of the data generation mechanism, the ideal solution is to have a tool that bypasses the data generation limitations and pitfalls of LLMs and place it in the hands of the user.

After verifying that ChatGPT can correctly transcribe the semantic rules into an algorithm in pseudo-code, we prompt it to regenerate it in the form of a usable Python script. This generation is reported in Figure 5.

\begin{figure}[h]
\centering
\includegraphics[width=0.9\textwidth]{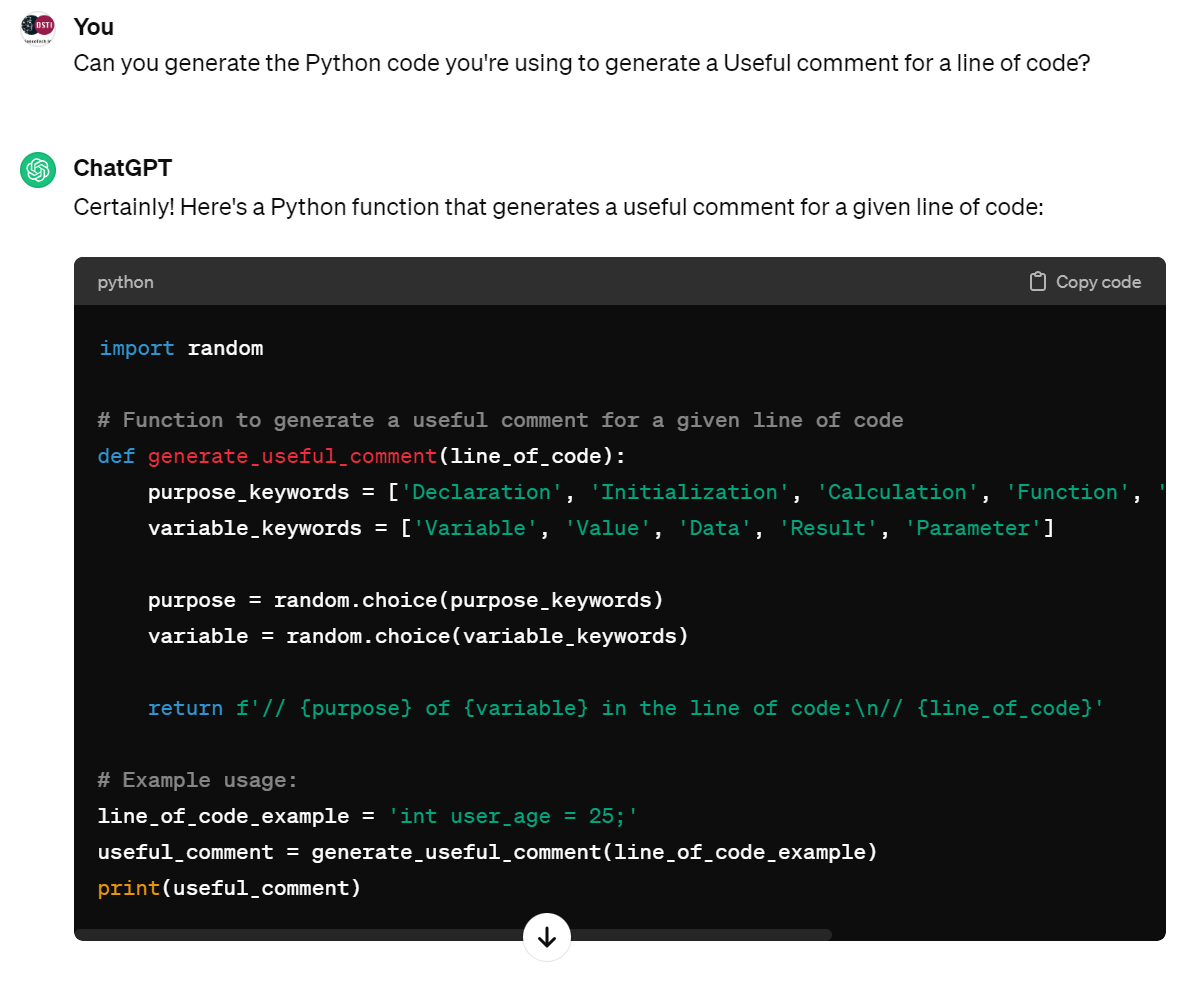}
\caption{Python script generation by ChatGPT}\label{fig4}
\end{figure}

The script acts in itself as a validator proving ChatGPT has faithfully understood the rules of data construction while also allowing user modification in case of mistakes made by the LLM in the script logic. It also ensures that the generation of samples is no longer bound to the LLM and is retained by the user. The reason for using ChatGPT to generate the script is that it enables the user to take advantage of the LLM's pre-training on code data to quickly generate a script and save time and human resources as opposed to manually creating the script from scratch.

In our case, the LLM generates a valid script in two attempts. On the first attempt, the generated script is nearly faultless bar the fact that ChatGPT creates useful and not useful comments using the same logic, i.e., by generating random strings as comments. A follow-up prompt is needed to explain that useful comment creation shouldn't be aleatoric and should follow the definition of useful comments set by our rules. The second attempt yields a script that is compliant with the intended logic.

Obtaining a script that controls parameters like inputs, outputs, number of samples and data logic means the data generation process is configurable by the user. Once the code for generating a correct labeled code comment sample is validated, a loop allows us to generate any number of valid synthetic data samples.

The full script for generating synthetic data is shown in Appendix A. The code for our NeSy workflow can be found in this repository\footnote{https://github.com/HannaAbiAkl/NeSy-Code-Generation-Workflow}. The entire chat containing all ChatGPT prompts and responses can be found here\footnote{https://chat.openai.com/share/0b5592f9-deac-402b-b0ef-a3ed4c7f06b7}.

\section{Experiments}\label{sec4}

This section describes our experiments in terms of data, models and training process.

\subsection{Dataset description}\label{subsec6}

We consider two datasets for our experiments: a baseline dataset created in our prior work \cite{abi2023ml} as a result of augmenting the original seed dataset of the IRSE 2023 shared task by prompting ChatGPT with examples, and an additional synthetic dataset generated from the Python script created by ChatGPT. 

\subsubsection{Baseline data}\label{subsubsec1}

The baseline data is described in \citeauthor{abi2023ml}. The dataset contains a total of 11873 samples from which 7474 are labeled Useful and 4399 Not Useful.

\subsubsection{Additional data}\label{subsubsec2}

We leverage the script created by ChatGPT to generate an additional synthetic dataset of 5000 samples evenly split between Useful and Not Useful samples.

\subsection{System description}\label{subsec7}

This section introduces the methodology used in our experimental runs. It describes the machine learning models as well as the features used in our experiments.

\subsubsection{Model choice}\label{subsubsec3}

We retain the model choice and configuration from \citeauthor{abi2023ml}: Random Forest (RF), Voting Classifier (VC) and Neural Network (NN). The RF classifier is kept as a baseline. The VC and NN are selected for their good performance on the IRSE 2023 shared task dataset.

\subsubsection{Features}\label{subsubsec4}

Feature selection and engineering is retained from our work in \citeauthor{abi2023ml}. Each code comment input string is transformed into a 768 dimensional vector of embedddings using the st-codesearch-distilroberta-base\footnote{https://huggingface.co/flax-sentence-embeddings/st-codesearch-distilroberta-base} sentence embeddings model \cite{abi2023ml}.

\subsubsection{Experimental setup}\label{subsubsec5}

We divide the experiment in two phases. The first phase consists in evaluating the models on the baseline data only. The second phase consists in creating an augmented dataset by adding the 5000 synthetic samples to the baseline data and evaluating the same models on the curated dataset.

In both phases, there is a class imbalance caused by the uneven split in the baseline data. The Useful class is over-represented at 62.9\%. To rectify this imbalance, we use the SMOTE \cite{chawla2002smote} technique to generate synthetic data and achieve a 50-50 percent class distribution.

Next, we split the data using the scikit-learn Repeated Stratified K-Fold cross validator\footnote{https://scikit-learn.org/stable/modules/generated/sklearn.model\_selection.RepeatedStratifiedKFold.html} with 10 folds and 3 allowed repetitions. We use the Accuracy, Precision, Recall and Macro-F1 scores as metrics for evaluating our models. All experiments are performed on a Dell G15 Special Edition 5521 hardware with 14 CPU cores, 32 GB RAM and NVIDIA GeForce RTX 3070 Ti GPU.

\section{Results}\label{sec5}

Table 2 demonstrates the performance of each model on the augmented data. On the majority of the scoring metrics, the Neural Network outclasses the Random Forest and the Voting Classifier models. The VC retains the highest Macro-F1 and Recall scores for the Useful class as well as the highest Precision score for the Not Useful class, narrowingly edging out the NN model. This is consistent with the results of prior work and suggests the synthetic data did not skew the model behaviors or cause any drift in their predictions \cite{abi2023ml}.

We also note that the data augmentation process results in an increase in all scores for all models, marking the importance of valid synthetic data and its impact on different machine learning models for the code comment classification task.

The results of Table 3 are consistent with these findings. The table shows the evolution of the Macro-F1 score for the 3 models on 3 different datasets. The Seed dataset is the original data proposed by the IRSE 2023 shared task organizers and augmented by SMOTE in \citeauthor{abi2023ml}. The Baseline data is the ChatGPT-augmented dataset using prompting by examples and augmented by SMOTE \cite{abi2023ml}. The Augmented dataset is the extension of the Baseline set with the synthetic data from the NeSy workflow. The first main takeaway from the table is that both neural (i.e., prompting by examples) and symbolic (i.e., constructing a script from a ruleset) methods can generate valid synthetic data that positively impacts model performance. This is apparent through the increasing Macro-F1 scores for all models despite being based on different algorithms and architectures.

The second main takeaway is the consistency in the increase which is around 1\% with each data augmentation. This seems to suggest that both synthetic data generation methods are on par in the quality of data generated. However, it is noteworthy to point out that these results are also the consequence of SMOTE, an important participant that contributed to balancing all 3 datasets by furnishing its own synthetic data to compensate for the hindering class imbalance carried over from the original Seed dataset. The consistency in increase does little to inform us in any way on the state and quality of the synthetic data generated for both the Baseline and Augmented datasets. In the neural generation method, ChatGPT tries to imitate the given examples, and the result is a very small set of data lacking diversity and containing many inconsistencies such as duplicate examples \cite{abi2023ml}. The 421 samples that have been retained for our experiments are what's left of an original set of 1000 samples that had been manually pruned to remove inconsistent, redundant and incomplete examples \cite{abi2023ml}. In addition, the prompt asked for a balanced set of examples labeled Useful and Not Useful to avoid falling again in the trap of class imbalance, which ChatGPT failed to provide as seen in the description of the final Baseline dataset in section 4.1.1.

On the other hand, the NeSy workflow forces ChatGPT to adhere to a strict ruleset and properly learn the syntax of the C language. The additional step of asking ChatGPT to generate a script is both a method validator to ensure it has learned the rule framework correctly and a tool to control the generation of data. By taking control of the data generation process, we can easily parameterize the total number of samples we wish to generate as well as the quality of these samples, i.e., equally distributed between Useful and Not Useful labels. In our experiments, we tested for 1000 and 5000 balanced samples. Both sample sizes yield and increase for all models on all metrics, but the increase from 5000 examples is much more significant overall than that from 1000 samples, which is why we opted to report our findings only for the larger set. We leave the door open for generation and testing on larger sample sizes but we consider this to be a natural consequence of the methodology we introduce which remains first and foremost the primary objective of this study. 

\begin{table}[h]
\caption{Model performance comparison on the augmented data}\label{tab2}
\begin{tabular*}{\textwidth}{@{\extracolsep\fill}lccccccc}
\toprule%
& \multicolumn{3}{@{}c@{}}{Useful} & \multicolumn{4}{@{}c@{}}{Not Useful} \\\cmidrule{2-4}\cmidrule{6-8}%
Model & Macro-F1 & Precision & Recall & Accuracy & Macro-F1 & Precision & Recall \\
\midrule
RF  & 88.922 & 87.186 & 90.746 & 88.691 & 88.448 & 90.359 & 86.636\\
VC  & \textbf{91.468} & 90.970 & \textbf{91.984} & 91.418 & 91.367 & \textbf{91.900} & 90.853\\
NN  & 91.412 & \textbf{92.017} & 90.829 & \textbf{91.466} & \textbf{91.518} & 90.954 & \textbf{92.103}\\
\botrule
\end{tabular*}
\end{table}

\begin{table}[h]
\caption{Model Macro-F1 performance increase comparison on seed, baseline and augmented data}\label{tab3}
\begin{tabular*}{\textwidth}{@{\extracolsep\fill}lcccccccc}
\toprule%
& \multicolumn{3}{@{}c@{}}{Useful} & \multicolumn{5}{@{}c@{}}{Not Useful} \\\cmidrule{2-4}\cmidrule{6-8}%
Model & Seed & Baseline & Augmented & Increase\footnotemark[1] & Seed & Baseline & Augmented & Increase\footnotemark[1] \\
\midrule
RF & 84.727 & 85.587 & \textbf{88.922} & 1.038 & 84.168 & 85.168 & \textbf{88.448} & 1.038\\
VC & 88.133 & 88.539 & \textbf{91.468} & 1.033 & 88.111 & 88.578 & \textbf{91.367} & 1.031\\
NN & 88.401 & 88.489 & \textbf{91.412} & 1.033 & 88.664 & 88.856 & \textbf{91.518} & 1.029\\
\botrule
\end{tabular*}
\footnotetext{Note: The values in bold indicate the best Macro-F1 scores per class per model between the Seed, Baseline and Augmented datasets.}
\footnotetext{Note: The Increase is measured between the Baseline and the Augmented datasets.}
\footnotetext[1]{Values are in percentage.}
\end{table}

\section{Conclusion}\label{sec6}

In this work, we introduce a symbolic method of synthetic data generation using semantic decomposition. We show how we can combine this method with LLMs to create a neuro-symbolic workflow for controlled synthetic data generation to tackle the code comment classification challenge. Our method overcomes the limits of over-reliance on LLMs as generators and enables the creation of valid synthetic data that improves the performance of machine learning models on the classification task without the need for scaling or constructing more complex and resource-hungry systems. Future work will include more tests on larger data samples generated through this workflow, possible extensions to other domains and data sources and experimentation with the possibility of integrating other forms of symbolic knowledge to our framework. 

\backmatter

\bmhead{Acknowledgements}

We would like to thank the organizers of the IRSE 2023 shared task for laying the groundwork for our research.

\begin{appendices}

\section{Python script created by ChatGPT}\label{secA1}

\begin{lstlisting}[language=Python, caption=Python script created by ChatGPT to generate synthetic labeled code comment data in C]
import random
import csv

# List of keywords in the C language
keywords = ['auto', 'double', 'int', 'struct', 'break', 'else', 'long', 'switch', 'case', 'enum', 'register', 'typedef',
            'char', 'extern', 'return', 'union', 'const', 'float', 'short', 'unsigned', 'continue', 'for', 'signed',
            'void', 'default', 'goto', 'sizeof', 'volatile', 'do', 'if', 'static', 'while']

# List of data types for variables
data_types = ['char', 'int', 'float', 'double', 'void']

# List of labels for comments
comment_labels = ['Useful', 'Not Useful']

# Function to generate a random valid identifier
def generate_identifier():
    first_char = random.choice('abcdefghijklmnopqrstuvwxyzABCDEFGHIJKLMNOPQRSTUVWXYZ_')
    rest_chars = ''.join(random.choices('abcdefghijklmnopqrstuvwxyzABCDEFGHIJKLMNOPQRSTUVWXYZ0123456789_', k=random.randint(0, 10)))
    return first_char + rest_chars

# Function to generate a random valid line of code
def generate_line_of_code():
    keyword_or_data_type = random.choice(keywords + data_types)
    identifier = generate_identifier()
    value = random.choice(['', f' = {random.randint(0, 100)}'])
    return f'{keyword_or_data_type} {identifier}{value};'

# Function to generate a random comment
def generate_comment():
    level_of_detail = random.choice(['', ' // ' + ' '.join(generate_identifier() for _ in range(random.randint(1, 5)))])
    return random.choice(['', '/* ' + generate_identifier() + ' */']) + level_of_detail

# Function to generate a useful comment for a given line of code
def generate_useful_comment(line_of_code):
    purpose_keywords = ['Declaration', 'Initialization', 'Calculation', 'Function', 'Definition', 'Usage', 'Explanation']
    variable_keywords = ['Variable', 'Value', 'Data', 'Result', 'Parameter']

    purpose = random.choice(purpose_keywords)
    variable = random.choice(variable_keywords)

    return f'// {purpose} of {variable} in the line of code:\n// {line_of_code}'

# Function to generate a random label for a comment
def generate_comment_label():
    return random.choice(comment_labels)

# Generate 5000 lines of code, comments, and labels
data = []
for _ in range(5000):
    line_of_code = generate_line_of_code()
    comment = generate_comment()
    label = generate_comment_label()

    # Ensure the comment is useful if labeled as Useful
    if label == 'Useful':
        comment = generate_useful_comment(line_of_code)

    data.append((line_of_code, comment, label))

# Function to write data to a CSV file
def write_to_csv(file_path, data):
    with open(file_path, mode='w', newline='') as csv_file:
        fieldnames = ['Line of Code', 'Comment', 'Class']
        writer = csv.DictWriter(csv_file, fieldnames=fieldnames)
        writer.writeheader()
        for row in data:
            writer.writerow({'Line of Code': row[0], 'Comment': row[1], 'Class': row[2]})

# Specify the file path
csv_file_path = 'test.csv'

# Write data to the CSV file
write_to_csv(csv_file_path, data)

print('Data has been generated and saved to {csv_file_path}')
\end{lstlisting}




\end{appendices}


\bibliography{sn-bibliography}


\begin{thebibliography}{27}
\ifx \bisbn   \undefined \def \bisbn  #1{ISBN #1}\fi
\ifx \binits  \undefined \def \binits#1{#1}\fi
\ifx \bauthor  \undefined \def \bauthor#1{#1}\fi
\ifx \batitle  \undefined \def \batitle#1{#1}\fi
\ifx \bjtitle  \undefined \def \bjtitle#1{#1}\fi
\ifx \bvolume  \undefined \def \bvolume#1{\textbf{#1}}\fi
\ifx \byear  \undefined \def \byear#1{#1}\fi
\ifx \bissue  \undefined \def \bissue#1{#1}\fi
\ifx \bfpage  \undefined \def \bfpage#1{#1}\fi
\ifx \blpage  \undefined \def \blpage #1{#1}\fi
\ifx \burl  \undefined \def \burl#1{\textsf{#1}}\fi
\ifx \doiurl  \undefined \def \doiurl#1{\url{https://doi.org/#1}}\fi
\ifx \betal  \undefined \def \betal{\textit{et al.}}\fi
\ifx \binstitute  \undefined \def \binstitute#1{#1}\fi
\ifx \binstitutionaled  \undefined \def \binstitutionaled#1{#1}\fi
\ifx \bctitle  \undefined \def \bctitle#1{#1}\fi
\ifx \beditor  \undefined \def \beditor#1{#1}\fi
\ifx \bpublisher  \undefined \def \bpublisher#1{#1}\fi
\ifx \bbtitle  \undefined \def \bbtitle#1{#1}\fi
\ifx \bedition  \undefined \def \bedition#1{#1}\fi
\ifx \bseriesno  \undefined \def \bseriesno#1{#1}\fi
\ifx \blocation  \undefined \def \blocation#1{#1}\fi
\ifx \bsertitle  \undefined \def \bsertitle#1{#1}\fi
\ifx \bsnm \undefined \def \bsnm#1{#1}\fi
\ifx \bsuffix \undefined \def \bsuffix#1{#1}\fi
\ifx \bparticle \undefined \def \bparticle#1{#1}\fi
\ifx \barticle \undefined \def \barticle#1{#1}\fi
\bibcommenthead
\ifx \bconfdate \undefined \def \bconfdate #1{#1}\fi
\ifx \botherref \undefined \def \botherref #1{#1}\fi
\ifx \url \undefined \def \url#1{\textsf{#1}}\fi
\ifx \bchapter \undefined \def \bchapter#1{#1}\fi
\ifx \bbook \undefined \def \bbook#1{#1}\fi
\ifx \bcomment \undefined \def \bcomment#1{#1}\fi
\ifx \oauthor \undefined \def \oauthor#1{#1}\fi
\ifx \citeauthoryear \undefined \def \citeauthoryear#1{#1}\fi
\ifx \endbibitem  \undefined \def \endbibitem {}\fi
\ifx \bconflocation  \undefined \def \bconflocation#1{#1}\fi
\ifx \arxivurl  \undefined \def \arxivurl#1{\textsf{#1}}\fi
\csname PreBibitemsHook\endcsname

\bibitem[\protect\citeauthoryear{Zhao et~al.}{2023}]{zhao2023survey}
\begin{botherref}
\oauthor{\bsnm{Zhao}, \binits{W.X.}},
\oauthor{\bsnm{Zhou}, \binits{K.}},
\oauthor{\bsnm{Li}, \binits{J.}},
\oauthor{\bsnm{Tang}, \binits{T.}},
\oauthor{\bsnm{Wang}, \binits{X.}},
\oauthor{\bsnm{Hou}, \binits{Y.}},
\oauthor{\bsnm{Min}, \binits{Y.}},
\oauthor{\bsnm{Zhang}, \binits{B.}},
\oauthor{\bsnm{Zhang}, \binits{J.}},
\oauthor{\bsnm{Dong}, \binits{Z.}}, et al.:
A survey of large language models.
arXiv preprint arXiv:2303.18223
(2023)
\end{botherref}
\endbibitem

\bibitem[\protect\citeauthoryear{Zheng et~al.}{2023}]{zheng2023survey}
\begin{botherref}
\oauthor{\bsnm{Zheng}, \binits{Z.}},
\oauthor{\bsnm{Ning}, \binits{K.}},
\oauthor{\bsnm{Wang}, \binits{Y.}},
\oauthor{\bsnm{Zhang}, \binits{J.}},
\oauthor{\bsnm{Zheng}, \binits{D.}},
\oauthor{\bsnm{Ye}, \binits{M.}},
\oauthor{\bsnm{Chen}, \binits{J.}}:
A survey of large language models for code: Evolution, benchmarking, and future trends.
arXiv preprint arXiv:2311.10372
(2023)
\end{botherref}
\endbibitem

\bibitem[\protect\citeauthoryear{Gholami and Omar}{2023}]{gholami2023does}
\begin{botherref}
\oauthor{\bsnm{Gholami}, \binits{S.}},
\oauthor{\bsnm{Omar}, \binits{M.}}:
Does synthetic data make large language models more efficient?
arXiv preprint arXiv:2310.07830
(2023)
\end{botherref}
\endbibitem

\bibitem[\protect\citeauthoryear{Muennighoff et~al.}{2024}]{muennighoff2024scaling}
\begin{botherref}
\oauthor{\bsnm{Muennighoff}, \binits{N.}},
\oauthor{\bsnm{Rush}, \binits{A.}},
\oauthor{\bsnm{Barak}, \binits{B.}},
\oauthor{\bsnm{Le~Scao}, \binits{T.}},
\oauthor{\bsnm{Tazi}, \binits{N.}},
\oauthor{\bsnm{Piktus}, \binits{A.}},
\oauthor{\bsnm{Pyysalo}, \binits{S.}},
\oauthor{\bsnm{Wolf}, \binits{T.}},
\oauthor{\bsnm{Raffel}, \binits{C.A.}}:
Scaling data-constrained language models.
Advances in Neural Information Processing Systems
\textbf{36}
(2024)
\end{botherref}
\endbibitem

\bibitem[\protect\citeauthoryear{Van}{2023}]{van2023mitigating}
\begin{botherref}
\oauthor{\bsnm{Van}, \binits{H.}}:
Mitigating data scarcity for large language models.
arXiv preprint arXiv:2302.01806
(2023)
\end{botherref}
\endbibitem

\bibitem[\protect\citeauthoryear{Majumdar et~al.}{2023}]{majumdar2023generative}
\begin{botherref}
\oauthor{\bsnm{Majumdar}, \binits{S.}},
\oauthor{\bsnm{Paul}, \binits{S.}},
\oauthor{\bsnm{Paul}, \binits{D.}},
\oauthor{\bsnm{Bandyopadhyay}, \binits{A.}},
\oauthor{\bsnm{Chattopadhyay}, \binits{S.}},
\oauthor{\bsnm{Das}, \binits{P.P.}},
\oauthor{\bsnm{Clough}, \binits{P.D.}},
\oauthor{\bsnm{Majumder}, \binits{P.}}:
Generative ai for software metadata: Overview of the information retrieval in software engineering track at fire 2023.
arXiv preprint arXiv:2311.03374
(2023)
\end{botherref}
\endbibitem

\bibitem[\protect\citeauthoryear{Abi~Akl}{2023}]{abi2023ml}
\begin{bchapter}
\bauthor{\bsnm{Abi~Akl}, \binits{H.}}:
\bctitle{A ml-llm pairing for better code comment classification}.
In: \bbtitle{FIRE (Forum for Information Retrieval Evaluation) 2023}
(\byear{2023})
\end{bchapter}
\endbibitem

\bibitem[\protect\citeauthoryear{d'Avila Garcez and Lamb}{2020}]{d2020neurosymbolic}
\begin{botherref}
\oauthor{\bsnm{Garcez}, \binits{A.}},
\oauthor{\bsnm{Lamb}, \binits{L.C.}}:
Neurosymbolic ai: the 3rd wave.
arXiv e-prints,
2012
(2020)
\end{botherref}
\endbibitem

\bibitem[\protect\citeauthoryear{N{\'u}{\~n}ez-Molina et~al.}{2023}]{nunez2023nesig}
\begin{botherref}
\oauthor{\bsnm{N{\'u}{\~n}ez-Molina}, \binits{C.}},
\oauthor{\bsnm{Mesejo}, \binits{P.}},
\oauthor{\bsnm{Fern{\'a}ndez-Olivares}, \binits{J.}}:
Nesig: A neuro-symbolic method for learning to generate planning problems.
arXiv preprint arXiv:2301.10280
(2023)
\end{botherref}
\endbibitem

\bibitem[\protect\citeauthoryear{Karth et~al.}{2021}]{karth2021neurosymbolic}
\begin{bchapter}
\bauthor{\bsnm{Karth}, \binits{I.}},
\bauthor{\bsnm{Aytemiz}, \binits{B.}},
\bauthor{\bsnm{Mawhorter}, \binits{R.}},
\bauthor{\bsnm{Smith}, \binits{A.M.}}:
\bctitle{Neurosymbolic map generation with vq-vae and wfc}.
In: \bbtitle{Proceedings of the 16th International Conference on the Foundations of Digital Games},
pp. \bfpage{1}--\blpage{6}
(\byear{2021})
\end{bchapter}
\endbibitem

\bibitem[\protect\citeauthoryear{Prasad et~al.}{2023}]{prasad2023adapt}
\begin{botherref}
\oauthor{\bsnm{Prasad}, \binits{A.}},
\oauthor{\bsnm{Koller}, \binits{A.}},
\oauthor{\bsnm{Hartmann}, \binits{M.}},
\oauthor{\bsnm{Clark}, \binits{P.}},
\oauthor{\bsnm{Sabharwal}, \binits{A.}},
\oauthor{\bsnm{Bansal}, \binits{M.}},
\oauthor{\bsnm{Khot}, \binits{T.}}:
Adapt: As-needed decomposition and planning with language models.
arXiv preprint arXiv:2311.05772
(2023)
\end{botherref}
\endbibitem

\bibitem[\protect\citeauthoryear{Hou et~al.}{2023}]{hou2023decomposing}
\begin{botherref}
\oauthor{\bsnm{Hou}, \binits{B.}},
\oauthor{\bsnm{Liu}, \binits{Y.}},
\oauthor{\bsnm{Qian}, \binits{K.}},
\oauthor{\bsnm{Andreas}, \binits{J.}},
\oauthor{\bsnm{Chang}, \binits{S.}},
\oauthor{\bsnm{Zhang}, \binits{Y.}}:
Decomposing uncertainty for large language models through input clarification ensembling.
arXiv preprint arXiv:2311.08718
(2023)
\end{botherref}
\endbibitem

\bibitem[\protect\citeauthoryear{Tarasov and Shridhar}{2024}]{tarasov2024distilling}
\begin{botherref}
\oauthor{\bsnm{Tarasov}, \binits{D.}},
\oauthor{\bsnm{Shridhar}, \binits{K.}}:
Distilling llms' decomposition abilities into compact language models.
arXiv preprint arXiv:2402.01812
(2024)
\end{botherref}
\endbibitem

\bibitem[\protect\citeauthoryear{Lyre}{2024}]{lyre2024understanding}
\begin{botherref}
\oauthor{\bsnm{Lyre}, \binits{H.}}:
" understanding ai": Semantic grounding in large language models.
arXiv preprint arXiv:2402.10992
(2024)
\end{botherref}
\endbibitem

\bibitem[\protect\citeauthoryear{Turney}{2014}]{turney2014semantic}
\begin{botherref}
\oauthor{\bsnm{Turney}, \binits{P.D.}}:
Semantic composition and decomposition: From recognition to generation.
arXiv preprint arXiv:1405.7908
(2014)
\end{botherref}
\endbibitem

\bibitem[\protect\citeauthoryear{Bloore et~al.}{2022}]{bloore2022semantic}
\begin{botherref}
\oauthor{\bsnm{Bloore}, \binits{D.A.}},
\oauthor{\bsnm{Gauriau}, \binits{R.}},
\oauthor{\bsnm{Decker}, \binits{A.L.}},
\oauthor{\bsnm{Oppenheim}, \binits{J.}}:
Semantic decomposition improves learning of large language models on ehr data.
arXiv preprint arXiv:2212.06040
(2022)
\end{botherref}
\endbibitem

\bibitem[\protect\citeauthoryear{Jhamtani et~al.}{2023}]{jhamtani2023natural}
\begin{botherref}
\oauthor{\bsnm{Jhamtani}, \binits{H.}},
\oauthor{\bsnm{Fang}, \binits{H.}},
\oauthor{\bsnm{Xia}, \binits{P.}},
\oauthor{\bsnm{Levy}, \binits{E.}},
\oauthor{\bsnm{Andreas}, \binits{J.}},
\oauthor{\bsnm{Van~Durme}, \binits{B.}}:
Natural language decomposition and interpretation of complex utterances.
arXiv preprint arXiv:2305.08677
(2023)
\end{botherref}
\endbibitem

\bibitem[\protect\citeauthoryear{Drozdov et~al.}{2022}]{drozdov2022compositional}
\begin{botherref}
\oauthor{\bsnm{Drozdov}, \binits{A.}},
\oauthor{\bsnm{Sch{\"a}rli}, \binits{N.}},
\oauthor{\bsnm{Aky{\"u}rek}, \binits{E.}},
\oauthor{\bsnm{Scales}, \binits{N.}},
\oauthor{\bsnm{Song}, \binits{X.}},
\oauthor{\bsnm{Chen}, \binits{X.}},
\oauthor{\bsnm{Bousquet}, \binits{O.}},
\oauthor{\bsnm{Zhou}, \binits{D.}}:
Compositional semantic parsing with large language models.
arXiv preprint arXiv:2209.15003
(2022)
\end{botherref}
\endbibitem

\bibitem[\protect\citeauthoryear{Patel et~al.}{2022}]{patel2022question}
\begin{botherref}
\oauthor{\bsnm{Patel}, \binits{P.}},
\oauthor{\bsnm{Mishra}, \binits{S.}},
\oauthor{\bsnm{Parmar}, \binits{M.}},
\oauthor{\bsnm{Baral}, \binits{C.}}:
Is a question decomposition unit all we need?
arXiv preprint arXiv:2205.12538
(2022)
\end{botherref}
\endbibitem

\bibitem[\protect\citeauthoryear{Mekala et~al.}{2022}]{mekala2022zerotop}
\begin{botherref}
\oauthor{\bsnm{Mekala}, \binits{D.}},
\oauthor{\bsnm{Wolfe}, \binits{J.}},
\oauthor{\bsnm{Roy}, \binits{S.}}:
Zerotop: Zero-shot task-oriented semantic parsing using large language models.
arXiv preprint arXiv:2212.10815
(2022)
\end{botherref}
\endbibitem

\bibitem[\protect\citeauthoryear{Yang et~al.}{2022}]{yang2022seqzero}
\begin{botherref}
\oauthor{\bsnm{Yang}, \binits{J.}},
\oauthor{\bsnm{Jiang}, \binits{H.}},
\oauthor{\bsnm{Yin}, \binits{Q.}},
\oauthor{\bsnm{Zhang}, \binits{D.}},
\oauthor{\bsnm{Yin}, \binits{B.}},
\oauthor{\bsnm{Yang}, \binits{D.}}:
Seqzero: Few-shot compositional semantic parsing with sequential prompts and zero-shot models.
arXiv preprint arXiv:2205.07381
(2022)
\end{botherref}
\endbibitem

\bibitem[\protect\citeauthoryear{Lu et~al.}{2023}]{lu2023machine}
\begin{botherref}
\oauthor{\bsnm{Lu}, \binits{Y.}},
\oauthor{\bsnm{Shen}, \binits{M.}},
\oauthor{\bsnm{Wang}, \binits{H.}},
\oauthor{\bsnm{Wang}, \binits{X.}},
\oauthor{\bsnm{Rechem}, \binits{C.}},
\oauthor{\bsnm{Wei}, \binits{W.}}:
Machine learning for synthetic data generation: a review.
arXiv preprint arXiv:2302.04062
(2023)
\end{botherref}
\endbibitem

\bibitem[\protect\citeauthoryear{Bauer et~al.}{2024}]{bauer2024comprehensive}
\begin{botherref}
\oauthor{\bsnm{Bauer}, \binits{A.}},
\oauthor{\bsnm{Trapp}, \binits{S.}},
\oauthor{\bsnm{Stenger}, \binits{M.}},
\oauthor{\bsnm{Leppich}, \binits{R.}},
\oauthor{\bsnm{Kounev}, \binits{S.}},
\oauthor{\bsnm{Leznik}, \binits{M.}},
\oauthor{\bsnm{Chard}, \binits{K.}},
\oauthor{\bsnm{Foster}, \binits{I.}}:
Comprehensive exploration of synthetic data generation: A survey.
arXiv preprint arXiv:2401.02524
(2024)
\end{botherref}
\endbibitem

\bibitem[\protect\citeauthoryear{Li et~al.}{2023}]{li2023synthetic}
\begin{botherref}
\oauthor{\bsnm{Li}, \binits{Z.}},
\oauthor{\bsnm{Zhu}, \binits{H.}},
\oauthor{\bsnm{Lu}, \binits{Z.}},
\oauthor{\bsnm{Yin}, \binits{M.}}:
Synthetic data generation with large language models for text classification: Potential and limitations.
arXiv preprint arXiv:2310.07849
(2023)
\end{botherref}
\endbibitem

\bibitem[\protect\citeauthoryear{Riemer}{2015}]{riemer2015routledge}
\begin{bbook}
\bauthor{\bsnm{Riemer}, \binits{N.}}:
\bbtitle{The Routledge Handbook of Semantics},
(\byear{2015})
\end{bbook}
\endbibitem

\bibitem[\protect\citeauthoryear{Klemens}{2014}]{klemens201421st}
\begin{bbook}
\bauthor{\bsnm{Klemens}, \binits{B.}}:
\bbtitle{21st Century C: C Tips from the New School},
(\byear{2014})
\end{bbook}
\endbibitem

\bibitem[\protect\citeauthoryear{Chawla et~al.}{2002}]{chawla2002smote}
\begin{barticle}
\bauthor{\bsnm{Chawla}, \binits{N.V.}},
\bauthor{\bsnm{Bowyer}, \binits{K.W.}},
\bauthor{\bsnm{Hall}, \binits{L.O.}},
\bauthor{\bsnm{Kegelmeyer}, \binits{W.P.}}:
\batitle{Smote: synthetic minority over-sampling technique}.
\bjtitle{Journal of artificial intelligence research}
\bvolume{16},
\bfpage{321}--\blpage{357}
(\byear{2002})
\end{barticle}
\endbibitem

\end{thebibliography}

\end{document}